\documentclass[twocolumn]{aastex62}
\usepackage{graphicx, bm, color, amsmath, amssymb, xfrac, accents, mathrsfs, comment, subfigure}

\def\be{\begin{equation}}
\def\ee{\end{equation}}
\newcommand{\f}{\frac}
\newcommand{\ac}{\accentset}
\newcommand{\ep}{\epsilon}
\newcommand{\om}{\omega}
\newcommand{\ph}{\phi_{\scriptscriptstyle\rm VEV}}
\newcommand{\Mpl}{M_{\scriptscriptstyle\rm Pl}}
\newcommand{\W}{\scriptscriptstyle\rm WD}
\newcommand{\N}{\scriptscriptstyle\rm NS}

\begin{document}

\title{Constraints of general screened modified gravities from comprehensive analysis of binary pulsars}

\author[0000-0001-5435-6502]{Xing Zhang}
\author{Wen Zhao}
\author{Tan Liu}
\affiliation{CAS Key Laboratory for Researches in Galaxies and Cosmology, Department of Astronomy, University of Science and Technology of China, Chinese Academy of Sciences, Hefei, Anhui 230026, China; starzhx@ustc.edu.cn, wzhao7@ustc.edu.cn}
\affiliation{School of Astronomy and Space Science, University of Science and Technology of China, Hefei 230026, China}

\author{Kai Lin}
\affiliation{Hubei Subsurface Multi-scale Imaging Key Laboratory, Institute of Geophysics and Geomatics, China University of Geosciences, Wuhan, Hubei, 430074, China}
\affiliation{Escola de Engenharia de Lorena, Universidade de S\~ao Paulo, 12602-810, Lorena, SP, Brazil}

\author{Chao Zhang}
\author{Xiang Zhao}
\affiliation{GCAP-CASPER, Physics Department, Baylor University, Waco, TX 76798-7316, USA}

\author{Shaojun Zhang}
\author{Tao Zhu}
\affiliation{Institute for Advanced Physics $\&$ Mathematics, Zhejiang University of Technology, Hangzhou 310032, China}

\author{Anzhong Wang}
\affiliation{GCAP-CASPER, Physics Department, Baylor University, Waco, TX 76798-7316, USA}
\affiliation{Institute for Advanced Physics $\&$ Mathematics, Zhejiang University of Technology, Hangzhou 310032, China}

\begin{abstract}
Testing gravity by binary pulsars nowadays becomes a key issue. Screened modified gravity is a kind of scalar-tensor theory with screening mechanism in order to satisfy the tight Solar System tests. In this paper, we investigate how the screening mechanism affects the orbital dynamics of binary pulsars, and calculate in detail the five post-Keplerian (PK) parameters in this theory. These parameters differ from those of general relativity (GR), and the differences are quantified by the scalar charges, which lead to the dipole radiation in this theory. We combine the observables of PK parameters for the ten binary pulsars, respectively, to place the constraints on the scalar charges and possible deviations from GR. The dipole radiation in the neutron star (NS) - white dwarf (WD) binaries leads to more stringent constraints on deviations from GR. The most constraining systems for the scalar charges of NS and WD are PSR~B1913$+$16 and PSR~J1738$+$0333, respectively. The results of all tests exclude significant strong-field deviations and show good agreement with GR.
\end{abstract}

\keywords{binaries: close -- gravitation -- gravitational waves -- pulsars: general}

\section{Introduction}
Pulsars are wonderful gravitational probes. Their environments involve strong gravitational fields, and they can provide us with much information about gravity. Since the first pulsar discovery in 1967~\citep{Hewish:1968aa}, more than 2600 pulsars have been discovered, and the binary pulsars comprise about 10\% of the known pulsar population\footnote{See the ATNF Pulsar Catalogue\\ {\tt http://www.atnf.csiro.au/people/pulsar/psrcat}}. The millisecond pulsars (MSPs) are pulsars with very short spin period (1$-$30 ms) \citep{Kramer:1998aa,Lorimer:2008aa}, and are often found in binary pulsar systems \citep{Bhattacharya:1991aa}. The spin period of MSPs is very stable and comparable to that of the best atomic clocks on Earth \citep{Taylor:1991aa, Hobbs:2012aa}. So great period stability makes them the wonderful probes of relativistic gravity. MSPs can be used to form a pulsar timing array to directly detect nanohertz gravitational waves (GWs) from astrophysical sources \citep{Sazhin:1978aa, Detweiler:1979aa, Yunes:2013aa}. Furthermore, the monitoring of MSPs allows us to perform high-precision pulsar timing \citep{Stairs:2003aa}, which allows us to directly measure not only the Keplerian parameters that define the characteristics of the Newtonian orbit, but also the relativistic corrections to the Keplerian orbit, which are quantified by the post-Keplerian (PK) parameters \citep{Damour:1986aa, Damour:1992aa}.

There are large number of relativistic parameters \citep{Damour:1992aa}, which can be obtained by analyzing the pulsar timing data. However, the most constrained are the five PK parameters: the periastron advance $\dot\omega$, the amplitude $\gamma$ of Einstein delay, the orbital period decay rate $\dot P_{b}$, the range $r$ and shape $s$ of Shapiro delay. In general relativity (GR), the PK parameters depend on only two unknown masses\footnote{In general, we do not consider the spin angular momentum, because the spin effects in binary pulsars are usually very small~\citep{Bhat:2008aa,Antoniadis:2013aa}.} of the pulsar and its companion, which is a consequence of the strong equivalence principle (SEP). However, it is not the case in most alternative theories of gravity, where the extra degrees of freedom affect the spacetime curvature and break the SEP \citep{Will:2014aa}. In alternative theories, the PK parameters are usually different from those of GR and play a crucial role in testing GR \citep{Taylor:1982aa,Taylor:1989aa,De-Laurentis:2012aa,De-Laurentis:2013aa}.

As the minimal extension of GR, the scalar-tensor gravity is a natural alternative \citep{Damour:1992ab,Fujii:2003aa,Capozziello:2011aa}, which invokes a conformal coupling between matter and scalar field. Meanwhile, in order to evade the tight Solar System tests, the screening mechanism is needed in this theory. This kind of scalar-tensor theory with screening mechanism can be described within a unified theoretical framework called screened modified gravity (SMG) \citep{Brax:2012aa}, which includes chameleon \citep{Khoury:2004aa}, symmetron \citep{Hinterbichler:2010es}, dilaton \citep{Damour:1994ab} and $f(R)$ \citep{Sotiriou:2010aa,De-Felice:2010aa} theories. In order to generate the screening mechanism in this theory, the effective potential of the scalar field must have a physical vacuum \citep{Brax:2012aa}. Around this vacuum, the scalar field can acquire an environment-dependent mass, which is required to increase as the ambient density increases. Therefore, the scalar force (i.e., fifth force) can be screened and evade the tight tests in dense regions (e.g., the Solar System), where the force range is so short that it is hard to detect within current experimental accuracy \citep{Gubser:2004aa}. Whereas in sparse regions (e.g., the Galaxy and the Universe), the long-range force may influence galactic dynamics \citep{Gronke:2015aa,Schmidt:2010aa}, and the scalar field can act as dark energy to accelerate the expansion of the Universe \citep{Khoury:2004aa,Hinterbichler:2011aa}. In addition, in this theory the tensor GWs have two polarization modes and all propagate with the speed of light \citep{Liu:2018ab}, whereby the severe constraints on the GWs speed from GW170817 are satisfied \citep{Abbott:2017ac,Abbott:2017aa}.

In this paper, we investigate how the screening mechanism in SMG affects the PK effects in binary pulsars, and discuss how to place the constraints on this theory by the pulsar observations. For our purposes, we study the impact of the scalar field on the orbital dynamics of binary pulsars and then calculate in detail the five PK parameters in this theory in the case of a quasi-elliptical orbit. These parameters depend not only on the masses of the pulsar and its companion but also on their scalar charges, which character the difference from GR. This theory can converge back to GR in the case of vanishing scalar charges. In this theory, the scalar charge of the object is inversely correlated to its compactness. Thus, for compact objects, the difference from GR is small and weak, which is completely different from other alternative theories without screening mechanisms.

In order to compare our results to the pulsar observations, we use the five neutron star (NS) - white dwarf (WD) binary pulsars and the five NS-NS binary pulsars. By combining the observables of PK parameters for the ten systems, respectively, performing Monte Carlo simulations, we place the constraints on the scalar charges of these systems. The dipole radiation in NS-WD binary pulsars dominates the orbital period decay, depends on the difference in the scalar charges, and leads to more stringent constraints on deviations from GR. We find that PSR~B1913$+$16 is the most constraining system for NS scalar charge. PSR~J1738$+$0333 is the most constraining system for WD scalar charge and provides the best upper bound on the scalar background of $3.4\times10^{-8}$ at 95.4\% confidence level (CL). In all tests from these systems we rule out significant strong-field deviations and find good agreement with GR.

This paper is organized as follows. In Section \ref{section2}, we display the action and field equations for SMG and perform post-Newtonian (PN) expansions of the tensor and scalar fields. In Section \ref{section3}, we calculate in detail the five PK parameters in this theory. In Section \ref{section4}, we discuss how to constrain this theory by the pulsar observations. In Section \ref{section5}, we perform Monte Carlo simulations for the ten binary pulsars, respectively, and discuss in detail these results. We conclude in Section \ref{section6} with a summary and discussion.

\section{Screened Modified Gravity}\label{section2}
\subsection{Action and Field Equations}
SMG is a kind of scalar-tensor theory with screening mechanism, which is characterized by the following action in the Einstein frame \citep{Brax:2012aa,Zhang:2017aa},
\begin{align}
\label{action_0}
\begin{split}
S=&\int d^4x\sqrt{-g}\left[\frac{\Mpl^2}{2}R-\frac12(\nabla\phi)^2-V(\phi)\right]
\\&+S_m\left[A^2(\phi) g_{\mu\nu},\,\psi_m^{(i)}\right],
\end{split}
\end{align}
where $\Mpl=\sqrt{1/8 \pi G}$ is the reduced Planck mass\footnote{In this paper we set the units to $c=\hbar=1$.}, $R$ is the Ricci scalar, and $\psi_m^{(i)}$ are various matter fields. The bare potential $V(\phi)$ can play the role of dark energy and endows the scalar field with mass. The coupling function $A(\phi)$ characterizes the interaction between matter and scalar field and induces the scalar force.

The scalar field is controlled by the effective potential $V_{\rm eff}(\phi)$ in Equation \eqref{scalar_eom}, which depends on two functions $V(\phi)$ and $A(\phi)$. For suitably chosen them, the effective potential can have a minimum (i.e., physical vacuum),
\begin{align}
\label{mass_eff}
\frac{{\rm d} V_{\rm eff}}{{\rm d}\phi}\bigg|_{\phi_{\rm min}}=0,\quad~    m^2_{\rm eff}\equiv \frac{{\rm d}^2 V_{\rm eff}}{{\rm d}\phi^2}\bigg|_{\phi_{\rm min}}>0.
\end{align}
Around this minimum, the scalar field acquires an effective mass, which is required to increase as the ambient density increases, 
\begin{align}
\f{{\rm d} m_{\rm eff}(\rho)}{{\rm d}\rho}>0.
\end{align}
Therefore, the scalar field can be screened in high density environments, in which the force range is so short that it is hard to find. So, SMG can generate the screening mechanism to suppress the scalar force in dense regions and pass the Solar System \citep{Zhang:2016aa}.

Let us now turn to the matter action for the compact objects. For the strongly self-gravitating objects, \citeauthor{Eardley:1975aa} first showed that so long as they are far enough from each other, their motions can be effectively described by point particles with the mass as a function of the scalar field,
\begin{align}
\label{matter_action}
\begin{split}
S_m=&-\sum_a\int m_a(\phi)d\tau_a.
\end{split}
\end{align}
where $m_a(\phi)$ is the $\phi$-dependent mass of the $a$-th point-particle.

Performing the variation of the full action for the compact objects with respect to the tensor field and the scalar field, yields the tensor field equation
\begin{align}
G_{\mu\nu}\label{tensor_eom}
= 8 \pi G \left(T_{\mu\nu}+T_{\phi\mu\nu}\right),
\end{align}
and the scalar field equation
\begin{align}
\square_g\phi\label{scalar_eom}
=\frac{\partial V_{\rm eff}(\phi)}{{\partial }\phi},
\end{align}
where $T_{\mu\nu}\equiv-(2/\sqrt{-g})\delta S_m/\delta g^{\mu\nu}$ is the energy-momentum tensor of matter, $\square_g{\equiv}g^{\mu\nu}\nabla_{\mu}\nabla_{\nu}$ is the curved space d'Alembertian, and $G_{\mu\nu}$ is the Einstein tensor. Here, the energy-momentum tensor of the scalar field is
\begin{align}\label{Tuv_phi}
T_{\phi\mu\nu}(\phi)=\partial_\mu\phi\partial_\nu\phi-g_{\mu\nu}\left[\frac{1}{2}(\partial\phi)^2+V(\phi)\right],
\end{align}
and the effective potential is given by
\be\label{Veff2}
V_{\text {eff}}(\phi) = V(\phi)+{\rho}A(\phi),
\ee
where $\rho$ is the local environment density of the scalar field.

\subsection{Weak-Field Limit}
In the weak-field limit, the fields $g_{\mu\nu}$ and $\phi$ are expanded around the backgrounds as follows:
\begin{align}\label{perturbations}
g_{\mu\nu}=\eta_{\mu\nu}+h_{\mu\nu},\qquad \phi=\ph+\varphi,
\end{align}
where $\eta_{\mu\nu}$ is the Minkowski background, and $\ph$ is the vacuum expectation value (VEV) of the scalar field (i.e., the scalar background).

The weak-field equations are given by \citep{Zhang:2017aa}
\begin{align}\label{linear_tensor_eq}
\square{h}_{\mu\nu}=-16\pi G\Big(T_{\mu\nu}-\f{1}{2}\eta_{\mu\nu}T\Big),
\end{align}
and
\begin{align}\label{linear_scalar_eq}
\left(\square-m^2_s\right)\varphi=-16\pi GS
\end{align}
where $\square\equiv\eta^{\mu\nu}\partial_{\mu}\partial_{\nu}$ is the flat-space d'Alembertian, and 
\begin{align}
m^2_{s}=\frac{{\rm d}^2 V}{\rm d\phi^2}\bigg|_{\ph}\!\!+\rho_b \frac{{\rm d}^2 A}{\rm d\phi^2}\bigg|_{\ph}
\end{align}
is the effective mass of the scalar field in a homogeneous background density $\rho_b$.
In the near zone, the source term $S$ turns into the PN expression \citep{Zhang:2017aa},
\begin{align}
\label{scalar_source1}
\begin{split}
S=&-\frac{1}{16\pi G}\Bigg\{\ph^{-1}\sum_as_am_a\delta^3\big(\mathbf{r}-\mathbf{r}_a(t)\big)\bigg[1-\frac12v_a^2
\\&-\frac12\ac{(2)}{h}_k^k+\f{s'_a}{s_a}\ph^{-1}\ac{(2)}{\varphi}\bigg]+\ac{(2)}{h}_{ij}\partial_i\partial_j\ac{(2)}{\varphi}\Bigg\}+\mathcal{O}(v^6),
\end{split}
\end{align}
where $v_a$ is the velocity of the $a$-th object, $m_a\equiv m_a(\ph)$ is its inertial mass at $\ph$, and
\begin{subequations}
\label{sensitivities_s}
\be
s_a\equiv\frac{\partial(\ln m_a(\phi))}{\partial(\ln \phi)}\bigg|_{\ph},
\ee
\be
s'_a\equiv s_a^2-s_a+\frac{\partial^2(\ln m_a(\phi))}{\partial(\ln \phi)^2}\bigg|_{\ph}
\ee
\end{subequations}
are respectively the first and second sensitivities \citep{Alsing:2012aa}, which characterize how the gravitational binding energy of an object responds to its motion relative to the additional fields.

\subsection{PN Solutions}
We now perform the PN expansions of the fields \citep{Will:1993aa,Will:2014aa}. This will allow us to derive the PK effects. The PN expansions of the tensor field have been given in the previous work \citep{Zhang:2017aa}, and are listed in Equations \eqref{PN_metric_scalar}.
Here, we use the PN approximation to solve the scalar field equation \eqref{linear_scalar_eq} in the near zone, where we can neglect the scalar field mass $m_s$ of cosmological scales and the potential $V(\phi)$ corresponding to the dark energy. The detailed derivations are given in Appendix \ref{appendix_a}. The PN expansions of the tensor and scalar fields are listed as follows:
\begin{subequations}\label{PN_metric_scalar}
\begin{align}
\begin{split}
g_{00}=&-1+2\sum_a\frac{Gm_a}{r_a}-2\bigg(\sum_{a}\frac{Gm_a}{r_a}\bigg)^2+3\sum_a\frac{Gm_av_a^2}{r_a}
\\&-2\sum_a\sum_{b\ne a}\frac{G^2m_am_b}{r_ar_{ab}}\left(1+\frac{1}{2}\ep_a\ep_b\right)+\mathcal{O}(v^6),\label{PN_tensor_00}
\end{split}
\end{align}
\begin{align}
\begin{split}
g_{0j}=&-\!\frac{7}{2}\!\sum_a\!\frac{Gm_av_a^j}{r_a}\!-\!\frac{1}{2}\!\sum_{a}\!\frac{Gm_a}{r_a^3}(\mathbf{r}_a\!\cdot\! \mathbf{v}_a)\!(r^j\!-\!r_a^j)\!+\!\mathcal{O}(v^5),\label{PN_tensor_0j}
\end{split}
\end{align}
\begin{align}
\begin{split}
g_{ij}=&\delta_{ij}\left(1+2\sum_a\frac{Gm_a}{r_a}\right)+\mathcal{O}(v^4),\label{PN_tensor_ij}
\end{split}
\end{align}
\begin{align}
\begin{split}
{\varphi}=&-\Mpl\sum_a\frac{Gm_a\ep_a}{r_a}\bigg[1-\f12v_a^2-\sum_{b\ne a}\f{Gm_b}{r_{ab}}
\\&
-\f{s'_a}{s_a}\f{\Mpl}{\ph}\sum_{b\ne a}\f{Gm_b\ep_b}{r_{ab}}+\f{r_a}{2}\f{\partial^2r_a}{\partial t^2}\bigg]+\mathcal{O}(v^6),\label{PN_scalar}
\end{split}
\end{align}
\end{subequations}
where $r_a=\left|\mathbf{r}-\mathbf{r}_a(t)\right|$ and $r_{ab}=\left|\mathbf{r}_a(t)-\mathbf{r}_b(t)\right|$. Here, the quantity $\ep_a$ is usually called the scalar charge (or screened parameter), given by \citep{Zhang:2016aa}
\begin{align}\label{epsilon_a}
\ep_a\equiv\frac{\ph-\phi_a}{\Mpl\Phi_{a}},
\end{align}
where $\Phi_a=Gm_a/R_a$ is the compactness (i.e., negative Newtonian gravitational potential at the surface) of the $a$-th object, and $\phi_a$ is the position of the effective potential minimum inside the $a$-th object. Obviously, the scalar charge is inversely proportional to the compactness, which agrees with the screening mechanisms. Note that in general $\phi_a$ is inversely correlated to the matter density of the local environment of the scalar field \citep{Zhang:2016aa}. Since the background matter density is always much less than that of compact stars, we have $\ph\gg\phi_{a}$, and the scalar charge reduces to $\ep_a=\ph/(\Mpl\Phi_a)$.

In addition, by comparing with the two scalar solutions obtained by means of different methods, we obtain a useful relation (see Appendix \ref{appendix_a}),
\begin{align}\label{s_a0}
s_a=\frac{\ph}{2\Mpl}\ep_a,
\end{align}
which means that the sensitivity is completely equivalent to the scalar charge in SMG. The scalar charge is inversely proportional to the compactness. Therefore, for compact objects, the sensitivity effects are very weak and the deviations from GR are very small. This is very different from other alternative theories without screening mechanisms, which generally predict the significant difference from GR for compact objects \citep{Will:1989aa}.

\section{Post-Keplerian Parameters}\label{section3}
In this section, we calculate in detail the five PK parameters in SMG for the binary pulsar moving on a quasi-elliptical orbit, which forms a major part of this paper.

\subsection{Periastron Advance}
The periastron advance is one of the four classic Solar System tests of GR, we now derive this effect in SMG for the binary pulsar. In this theory, the scalar field modifies the conservative orbital dynamics of the system, which can be derived by using the method of \citeauthor*{Einstein:1938aa} (EIH). From the matter action \eqref{matter_action}, the matter Lagrangian for the $a$-th object is given by
\begin{align}
\label{lagrangian}
L_a = m_a(\phi)\left(-g_{00}-2g_{0i}v_a^i-g_{ij}v_a^iv_a^j\right)^\f{1}{2}.
\end{align}
We follow the procedure detailed after Equation (11.90) of \citet{Will:1993aa} to derive the $N$-body Lagrangian. By substituting the PN expressions and the expansion of $m_a(\phi)$, making the terms in $L_a$ manifestly symmetric under interchange of all pairs of particles, we take one of each such term generated in $L_a$ and sum over $a$. Up to $\mathcal{O}(v^4)$, the final form of the EIH Lagrangian is given by
\begin{align}
\begin{split}
L_{\rm EIH}=&-\sum_am_a\!\left(\!1-\frac12v_a^2-\f18v_a^4\!\right)\!+\!\frac12\sum_{a}\sum_{b\ne a}\frac{Gm_am_b}{r_{ab}}
\\&\!\times\!\bigg[\,\mathscr{G}_{ab}\,+\,3\mathscr{B}_{ab}v_a^2\,-\,\f12(\,\mathscr{G}_{ab}+6\mathscr{B}_{ab}\,)(\,\mathbf{v}_a\cdot \mathbf{v}_b\,)
\\&-\f12\mathscr{G}_{ab}(\mathbf{n}_{ab}\!\cdot\! \mathbf{v}_a)(\mathbf{n}_{ab}\!\cdot\! \mathbf{v}_b)\!-\!\sum_{c \ne a}\!\frac{Gm_c}{r_{ac}}\mathscr{D}_{abc}\bigg],
\end{split}
\end{align}
where $\mathbf{n}_{ab}=({\mathbf{r}_a-\mathbf{r}_b})/{r_{ab}}$ is the unit direction vector from the $b$-th object to the $a$-th object, and we have defined
\begin{subequations}
\label{GBD}
\be
\mathscr{G}_{ab}=1+\f{1}{2}\ep_a\ep_b,
\ee
\be
\mathscr{B}_{ab}=1-\f{1}{6}\ep_a\ep_b,
\ee
\be
\mathscr{D}_{abc}=1+\f{1}{2}\ep_a\ep_b+\f{1}{2}\ep_a\ep_c+s^{\prime}_a\ep_b\ep_c\left(\f{\Mpl}{\ph}\right)^2.
\ee
\end{subequations}
In the above calculations, we have neglected the scalar field mass $m_s$, since the separation $r_{ab}$ in the near zone is always much less than $m^{-1}_s$ of cosmological scales.

Now let us specialize to a two-body system (labeled by 1 and 2 ) with the center of mass at the origin. Substituting the EIH Lagrangian into the Euler-Lagrange equation, yields the two-body equation of motion,
\begin{align}
\begin{split}\label{motion_eq}
\mathbf{a}_{1}\!=&\!-\!\frac{Gm_2\mathbf{n}_{12}}{r^2}\!\bigg[\!\mathscr{G}\!-\!\mathscr{G}\!v_{\!1}^2\!+\!\f{1}{2}\!(\mathscr{G}\!+\!3\mathscr{B})\!(\!\mathbf{v}_{\!2}\!-\!\mathbf{v}_{\!1}\!)^{\!2}\!\!-\!\f{3}{2}\mathscr{G}\!(\!\mathbf{v}_{\!2}\!\cdot\!\mathbf{n}_{12}\!)^{\!2}
\\&
-(3\mathscr{G}\mathscr{B}+\mathscr{D}_{122})\f{Gm_2}{r}\!-\!(\mathscr{G}^2+3\mathscr{G}\mathscr{B}+\mathscr{D}_{211})\f{Gm_1}{r}\bigg]
\\&
-\f{Gm_2}{r^2}(\mathbf{v}_2\!-\!\mathbf{v}_1)\big[(\mathscr{G}\!+\!3\mathscr{B})\mathbf{v}_1\!-\!3\mathscr{B}\mathbf{v}_2\big]\!\cdot\!\mathbf{n}_{12},
\end{split}
\end{align}
where we have defined $r\equiv r_{12}$, $\mathscr{G}\equiv\mathscr{G}_{12}$ and $\mathscr{B}\equiv\mathscr{B}_{12}$ for a two-body system. The equation of the body 2 can be obtained by exchanging all the particle labels $1\leftrightarrow 2$. To Newtonian order, this result satisfies the inverse-square law, which guarantees that the Kepler's third law 
\begin{align}\label{Kepler_3rd}
\Big(\f{2\pi}{P_b}\Big)^2a^3=\mathcal{G}m
\end{align}
holds in SMG theories. Here, $a$ and $P_b$ are the semimajor axis and the orbital period, $m$ is the total mass of the system, and $\mathcal{G}=G(1+\f12\ep_1\ep_2)$ is the effective gravitational constant between two compact objects.

By using the two-body equation of motion, employing the method of osculating elements \citep{Will:1993aa}, we obtain the periastron advance of the binary system,
\begin{align}
\begin{split}
\dot{\om}=&\f{6\pi{Gm}}{a(1-e^2)P_b}\mathscr{P_b}\mathscr{G}^{-1},
\end{split}
\end{align}
where $e$ is the orbital eccentricity, and we have defined
\begin{align}
\begin{split}
\mathscr{P_b}=\mathscr{G}\mathscr{B}+\f{1}{6}\mathscr{G}^2-\f{1}{6}\f{m_1\mathscr{D}_{211}+m_2\mathscr{D}_{122}}{m}.
\end{split}
\end{align}
It can be seen that the above expression reduces to the GR result in the limit of $\ep\rightarrow0$ \citep{Will:2014aa}.

\subsection{Einstein Delay}
In the binary pulsar system with an elliptical orbit, one has a changing time dilation of the radio pulses due to a variation in the pulsar velocity and a change of the distance between the pulsar and its companion. This time delay effect is so-called Einstein delay. The proper time at the pulsar's point of emission can be related to coordinate time by
\begin{align}
\begin{split}
{d{\tau}_p}=dt_p\bigg[-g_{\mu\nu}(x_p)\f{dx_p^{\mu}}{dt_p}\f{dx_p^{\nu}}{dt_p}\bigg]^{\f{1}{2}}.
\end{split}
\end{align}
By substituting the PN expressions in \eqref{PN_metric_scalar}, integrating this equation and dropping the constant contribution, and the result to first order can be written as 
\begin{align}\label{Time_Delay}
\begin{split}
{\tau}_p={t_p}-\gamma\sin{E},
\end{split}
\end{align}
and the amplitude of Einstein delay is
\begin{align}
\begin{split}
\gamma=\f{\mathcal{G}m_c}{a}\Big(1+\f{m_c}{m}\Big)\f{P_b}{2\pi}e,
\end{split}
\end{align}
where $E$ and $m_c$ are the eccentric anomaly and the companion mass, respectively. Using the Kepler's third law, we can rewrite the amplitude as
\begin{align}
\begin{split}
{\gamma}=e\f{P_b}{2\pi}\left(\f{2\pi{\mathcal{G}m}}{P_b}\right)^{2/3}\f{m_c}{m}\left(1+\frac{m_c}{m}\right),
\end{split}
\end{align}
which is identical to that of GR in the limit of $\ep\rightarrow0$ \citep{Will:2014aa}.

\subsection{Shapiro Delay}\label{Shapiro}
In the binary pulsar system with a sufficiently edge-on orbit, the propagation time delay suffered by the pulsar signal passing through the curved spacetime region near the companion, is so-called \citeauthor{Shapiro:1964aa} delay. It is also one of the four classic Solar System tests of GR.

This effect is described by two Shapiro delay terms called shape $s$ and range $r$ in the DD \citep{Damour:1986aa} parametrization, or other two quantities $\varsigma$ and $h_3$ in the FW \citep{Freire:2010aa} parametrization, which relate $s$ and $r$ by 
\begin{align}\label{Shapiro_delay0}
\begin{split}
s=\f{2\varsigma}{1+\varsigma^2},~~~~ r=\f{h_3}{\varsigma^3}.
\end{split}
\end{align}
In the DD formulation, to first order, the Shapiro delay is usually parameterised by \citep{Wex:2014aa}
\begin{align}\label{Shapiro_delay1}
\begin{split}
\bigtriangleup_{\rm S}=2r\ln\Big[1&-e\cos{E}-s\sin{\omega}(\cos{E}-e)
\\&-s\cos{\omega}(1-e^2)^{{1}/{2}}\sin{E}\Big],
\end{split}
\end{align}
where the quantities $\omega$, $e$ and $E$ are the longitude of periastron, the orbital eccentricity and the eccentric anomaly, respectively. The shape $s$ is linked to the orbit inclination $i$ by $s\equiv\sin{i}=x_p/a_p$, where $a_p$ and $x_p$ are the semimajor axis and projected semimajor axis of the pulsar orbit, respectively. Using this and the Kepler's third law, we obtain the Shapiro delay shape
\begin{align}
\begin{split}
s={x_p}\left(\f{2\pi}{{P_b}}\right)^{2/3}\f{m^{2/3}}{{\mathcal{G}^{1/3}}m_c}.
\end{split}
\end{align}

The pulsar signal travels along a null geodesic in the Jordan frame, $d{\tau_J}=A(\phi)d\tau=0$, that is
\begin{align}\label{null_geodesic_1}
\begin{split}
g_{\mu\nu}dx^{\mu}dx^{\nu}=0,
\end{split}
\end{align}
which indicates that the null geodesic remains unchanged under the conformal transformation. To order $\mathcal{O}(v^2)$, this can be written as
\begin{align}\label{null_geodesic_2}
\begin{split}
-1+\ac{(2)}h_{00}+(\delta_{ij}+\ac{(2)}h_{ij})\f{dx^i}{dt}\f{dx^j}{dt}=0.
\end{split}
\end{align}
The photon trajectory is corrected by $x^i(t)=x^i_p+n^i(t-t_p)+x^i_{\rm PN}(t)$, where the photon is emitted from $\mathbf{x}_p$ in direction $\mathbf{n}$ at time $t_p$, and $x^i_{\rm PN}$ is the PN correction. Substituting this and the PN expressions \eqref{PN_metric_scalar} into Equation \eqref{null_geodesic_2}, integrating this equation, we obtain the Shapiro delay
\begin{align}\label{Shapiro_delay2}
\begin{split}
\bigtriangleup_{\rm S}=2Gm_c\ln\bigg[\f{(r_p+\mathbf{r}_p\cdot\mathbf{n})(r_e+\mathbf{r}_e\cdot\mathbf{n})}{r_b^2}\bigg].
\end{split}
\end{align}
Here, the pulsar signal is emitted from the pulsar $\mathbf{r}_p$ in direction $\mathbf{n}$, passing through the curved spacetime region near the companion, and traveling to the Earth $\mathbf{r}_e$. The quantity $r_b$ is the impact parameter of the pulsar signal with respect to the companion. By comparing this result with Equation \eqref{Shapiro_delay1}, we obtain the Shapiro delay range
\begin{align}\label{Shapiro_r}
\begin{split}
r=Gm_c,
\end{split}
\end{align}
which is exactly the same as that in GR \citep{Will:2014aa}. This is because photons do not couple to the scalar field in this theory.

\subsection{Orbital Period Decay}
These four PK parameters calculated in the previous subsections all describe the effects of the conservative sector of the theory. The PK parameter $\dot{P_b}$ called the orbital period decay rate, is related to the damping of the orbital energy due to the emission of GWs, and characterizes the effects of the dissipative sector of the theory. In fact, it was the monitoring of the orbital period that led to the first indirect detection of GWs \citep{Hulse:1975aa,Taylor:1982aa,Taylor:1989aa}. In this subsection we calculate the orbital period decay rate $\dot{P_b}$ for the binary pulsar with a quasi-elliptical orbit.

In this theory, the tensor and scalar gravitational radiations contribute to the rate of the energy loss. The tensor energy flux can be obtain by performing the same way as GR \citep{Maggiore:2007aa}, and is given by (see \citet{Zhang:2017aa} for detailed derivations)
\begin{align}
\begin{split}\label{Quadrupole_tensor_flux}
\mathcal{F}_g^{Q}=\frac{32G^4\mu^2m^3}{5a^5}\f{1+\f{73}{24}e^2+\f{37}{96}e^4}{(1-e^2)^{7/2}}\big(1+\f{3}{2}\ep_p\ep_c\big),
\end{split}
\end{align}
where $m$ and $\mu$ are the total and reduced masses of the system, and $\ep_p$ and $\ep_c$ are the scalar charges of the pulsar and its companion.

The massive scalar wave equation \eqref{linear_scalar_eq} implies that the scalar particle satisfies the relativistic dispersion relation. This result shows that the scalar mode is excited if and only if its frequency (energy) is greater than its mass. For a typical binary pulsar with a 1-hour orbital period, the orbital frequency ($\om\sim10^{-3} {\rm Hz}$) is much greater than the scalar field mass $m_s$ of cosmological scales (if $m_s^{-1}\sim1{\rm Mpc}$, then $m_s\sim10^{-14}{\rm Hz}$). Because of this, we neglect the scalar field mass in the calculation of the scalar energy flux, which has been given in the recent work \citep{Zhang:2019aa},
\begin{widetext}
\begin{align}
\label{Dipole_scalar_flux_1}
\begin{split}
\mathcal{F}_{\phi}&=\frac{G^3\mu^2m^2}{6a^4}\f{(\ep_p-\ep_c)^2(1+\f{1}{2}e^2)}{(1-e^2)^{5/2}}+\f{G^4\mu^2m^3}{15a^5}\f{1}{(1-e^2)^{7/2}}\bigg\{\Big(8\ep_p\ep_c-5(\ep_p-\ep_c)^2\f{\mu}{m}\Big)+\f{e^2}{4}\Big[129\ep_p\ep_c-5(\ep_p-\ep_c)^2
\\&
+\!20\!\Big(\f{\ep_pm_p\!+\!\ep_cm_c}{m}\Big)^{\!2}\!+\!55\!\Big(\f{\ep_cm_p\!+\!\ep_pm_c}{m}\Big)^{\!2}\Big]\!+\!\f{e^4}{16}\!\Big[111\ep_p\ep_c\!+\!10(\ep_p\!-\!\ep_c)^{\!2}\!+\!5\!\Big(\f{\ep_pm_p\!+\!\ep_cm_c}{m}\Big)^{\!2}\!+\!40\!\Big(\f{\ep_cm_p\!+\!\ep_pm_c}{m}\Big)^{\!2}\Big]\!\bigg\}.
\end{split}
\end{align}
\end{widetext}
By summing this and the tensor energy flux \eqref{Quadrupole_tensor_flux}, using the relation 
\be
\f{\dot{P_b}}{P_b}=-\f{3}{2}\f{\dot{E}}{E},
\ee
the balance law
\be
\mathcal{F}=-\dot{E},
\ee
and the orbital binding energy 
\be
E=-\f{\mathcal{G}m\mu}{2a},
\ee
we obtain the orbital period decay rate due to the tensor and scalar radiations, presented in Equation \eqref{P_decay2}.

\subsection{Summary of PK parameters}
In the previous subsections, we have derived in detail the five PK parameters in this theory. For convenience, now we rewrite and summarize them in the following form:
\begin{subequations}\label{5_PK}
\begin{align}
\dot{\om}&=3\left(\f{P_{b}}{2\pi}\right)^{-\f{5}{3}}\f{(T_{\sun}m)^{\f{2}{3}}}{1-e^2}\f{24+8\ep_p\ep_c-\ep^2_p\ep^2_c}{24(1+\ep_p\ep_c/2)^{\f{4}{3}}},
\\
\gamma&=e\left(\f{P_b}{2\pi}\right)^{\f{1}{3}}T_{\sun}^{\f{2}{3}}\f{m_c}{m^{\f{1}{3}}}\Big(1+\f{m_c}{m}\Big)\Big(1+\f{1}{2}\ep_p\ep_c\Big)^{\f{2}{3}},
\\
r&=T_{\sun}{m_c},
\\
s&=x_p\left(\f{{P_b}}{2\pi}\right)^{-\f{2}{3}}T_{\sun}^{-\f{1}{3}}\f{m^{\f{2}{3}}}{m_c}\Big(1+\f{1}{2}\ep_p\ep_c\Big)^{-\f{1}{3}},
\\
\dot{P_b}&=\!-2\pi\left(\f{P_b}{2\pi}\right)^{\!\!-\f{5}{3}}\!\!\f{T_{\sun}^{\f{5}{3}}m_pm_c}{m^{\f13}(1-e^2)^{\f72}}
\bigg\{\f{96}{5}\Big(1\!+\!\f{73e^2}{24}\!+\!\f{37e^4}{96}\Big)
\nonumber\\&
+\!\left(\!\f{P_b}{2{\pi}T_{\sun}m}\!\right)^{\!\f23}\!\f{\ep_d^2}{2}\Big(1\!-\!\f{e^2}{2}\!-\!\f{e^4}{2}\Big)\!+\!\Big(8\ep_p\ep_c\!-\!\f{m_pm_c}{m^2}\ep_d^2\Big)
\nonumber\\&
+\f{e^2}{12}\Big[335\ep_p\ep_c+\Big(9-24\f{m_pm_c}{m^2}\Big)\ep_d^2+21\Gamma^2\Big]
\nonumber\\&
+\f{e^4}{48}\Big[191\ep_p\ep_c+\Big(9-6\f{m_pm_c}{m^2}\Big)\ep_d^2+21\Gamma^2\Big]
\bigg\}\label{P_decay2},
\end{align}
\end{subequations}
where masses are expressed in solar units, and $T_{\sun}\equiv G M_{\sun}= 4.925490947 \mu \rm s$ is a solar mass in time units. The quantity $\ep_d\equiv \ep_c-\ep_p$ is the difference in the scalar charges, and $\Gamma\equiv \ep_cm_p/m+\ep_pm_c/m$ is the mass-weighted average of the scalar charges. The PK parameters depend on the masses $m_p$ and $m_c$ and the scalar charges $\ep_p$ and $\ep_c$ of the pulsar and its companion, and can reduce to the results of GR in the limit of $\ep_p=\ep_c=0$ \citep{Will:2014aa}.

For the periastron advance, Einstein delay and  Shapiro delay, they are the effects of 1PN, 1PN and 1.5PN, respectively. In Equation \eqref{P_decay2}, the first and second terms are the tensor quadrupole radiation of 2.5 PN and the scalar dipole radiation of 1.5 PN, respectively, and the remaining terms represent the 2.5 PN contributions from the monopole and the monopole-quadrupole and dipole-octupole cross terms. In the case of $e=0$, $\dot{P_b}$ can reduce to the previous result \citep{Zhang:2017aa}. Because $P_b/T_{\sun}=\mathcal{O}(10^{9})$ for a typical binary pulsar with a 1-hour orbital period, the dipole radiation dominates the orbital period decay unless $\ep_c-\ep_p=0$.

\section{Binary pulsars}\label{section4}
In GR, the PK parameters depend on only two unknown masses of the pulsar and its companion. However, in SMG, the PK parameters \eqref{5_PK} contain the four unknown quantities, which are the masses and scalar charges of the pulsar and its companion. In this section, we discuss how to determine these unknown quantities by the observational data of the binary pulsar.

\subsection{Binary Pulsars}
Pulsars are extremely useful tools for testing GR and alternative theories of gravity, thanks to the extreme precision of their radio pulses. Pulsars have short spin period with a very stability. The monitoring of the times-of-arrival (TOAs) of the pulsar's radio pulses allows the properties of the pulsar orbit to be inferred. These orbital properties are parameterized by the Keplerian parameters and a suite of PK parameters, which describe the secular orbital changes over time as well as the propagation time delay. These orbital parameters can be precisely determined by using the standard software package {\sc tempo2} \citep{Hobbs:2006aa} to analyze the TOAs.

Note that, for the orbital period decay rate, in order to determine its intrinsic value caused by GWs damping, its measured value needs to be corrected for two effects. The first is the kinematic effect \citep{Damour:1991aa} caused by the relative acceleration between the pulsar system and the Solar System. The second is so-called \citeauthor{Shklovskii:1970aa} effect caused by the pulsar transverse motion relative to the Earth. The contributions of these effects can be removed by measuring the distance and proper motion of the pulsar system.

In this paper, we use the five NS-NS PSRs J0737$-$3039, B1534$+$12, J1756$-$2251, B1913$+$16 and J1757$-$1854, and the five NS-WD PSRs J1141$-$6545, J1738$+$0333, J0348$+$0432, J1012$+$5307 and J0751$+$1807. The reason for these choices is that they can provide at least three observables to test GR. For convenience, we list the parameters of these systems in Table \ref{tab1_PSRs_param}.

\begin{longrotatetable}
\begin{deluxetable*}{lllrrrrrrlll}
\tablecaption{Timing model parameters for the ten binary pulsars\label{tab1_PSRs_param}}
\tablewidth{0pt}
\tabletypesize{\scriptsize}
\tablehead{
\colhead{Parameters\tablenotemark{a}} & \colhead{$P_b$} & \colhead{$x$} & \colhead{$e$} & \colhead{$\dot\omega$} & \colhead{$\gamma$} & \colhead{$\dot P_b^{\text{int}}$} & \colhead{$s$ [or $\varsigma$]} & \colhead{$r$ [or $h_3$]} & \colhead{$q$} & \colhead{$m_c$} & \colhead{$m_p$}\\
\colhead{} & \colhead{(days)} & \colhead{(s)} & \colhead{} & \colhead{(deg/yr)} & \colhead{(ms)} & \colhead{$(10^{-12})$}& \colhead{} & \colhead{$(\mu s)$} & \colhead{} & \colhead{(M$_{\sun}$)} & \colhead{(M$_{\sun}$)}
} 
\startdata
{\bf NS$-$NS}\\
J0737$-$3039$^1$ &0.10225156248(5)& 1.415032(1)&0.0877775(9)& 16.89947(68)& 0.3856(26)& $-1.252(17)$& $0.99974^{+0.00016}_{-0.00039}$& 6.21(33) &1.0714(11)& $1.2489(7)$\tablenotemark{b}& $1.3381(7)$\tablenotemark{b}\\
B1534$+$12$^2$ &0.420737298879(2)& 3.7294636(6)& 0.27367752(7) & 1.7557950(19)& 2.0708(5)& \nodata& 0.9772(16) & 6.6(2)&\nodata&  $1.3455(2)$\tablenotemark{b}&  $1.3330(2)$\tablenotemark{b} \\
J1756$-$2251$^3$& 0.31963390143(3)& 2.756457(9)& 0.1805694(2) & 2.58240(4)& 1.148(9)& $-0.234^{+0.009}_{-0.006}$& 0.93(4)& 7.9(3.0)&\nodata & $1.230(7)$\tablenotemark{b}& $1.341(7)$\tablenotemark{b}\\
B1913$+$16$^4$ &0.322997448918(3)& 2.341776(2)& 0.6171340(4)& 4.226585(4)&4.307(4)&$-2.398(4)$& \nodata & \nodata&\nodata& $1.390(1)$\tablenotemark{b}& $1.438(1)$\tablenotemark{b}\\
J1757$-$1854$^5$ &0.18353783587(5)& 2.237805(5)& 0.6058142(10)& 10.3651(2)&3.587(12)&$-5.3(2)$& [0.90(3)] & [4.6(7)]&\nodata& $1.3946(9)$\tablenotemark{b}& $1.3384(9)$\tablenotemark{b}\\
{\bf NS$-$WD}\\
J1141$-$6545$^6$   & 0.1976509593(1)& 1.858922(6)& 0.171884(2)& 5.3096(4)& 0.773(11)&$-0.401(25)$& 0.97(1)&\nodata&\nodata& $1.02(1)$\tablenotemark{b}& $1.27(1)$\tablenotemark{b}\\
J1738$+$0333$^7$  &0.3547907398724(13)&0.343429130(17)& $0.34(11)\!\times\!10^{-6} $& \nodata& \nodata&$-0.0259(32)$& \nodata& \nodata&8.1(2)& $0.181^{+0.008}_{-0.007}$& $1.46^{+0.06}_{-0.05}$\\
J0348$+$0432$^8$ &0.102424062722(7)& 0.14097938(7)& $0.24(10)\!\times\!10^{-5} $& \nodata& \nodata &$-0.273(45)$  & \nodata& \nodata&11.70(13)& 0.172(3)& 2.01(4)\\
J1012$+$5307$^9$ & 0.60467271355(3)&0.5818172(2)&$1.2(3)\!\times\!10^{-6} $& \nodata& \nodata  & $ -0.029(21)$   & \nodata& \nodata &10.5(5)& 0.16(2)& 1.64(22) \\
J0751$+$1807$^{10}$ & 0.263144270792(7)& 0.3966158(3)&$3.3(5)\!\times\!10^{-6} $&\nodata& \nodata&$-0.0462(36)$& [0.81(17)]& [0.30(6)]&\nodata& 0.16(1)\tablenotemark{b}& $1.64(15)\tablenotemark{b}$\\
\enddata
\tablenotetext{a}{Orbital period $P_b$, projected semimajor axis $x$, eccentricity $e$, advance of periastron $\dot\omega$, amplitude of Einstein delay $\gamma$, intrinsic period derivative $\dot P_b^{\text{int}}$, ``shape'' $s$ [or $\varsigma$] and ``range'' $r$ [or $h_3$] of Shapiro delay, mass ratio $q=m_p/m_c$, companion mass $m_c$, pulsar mass $m_p$.}
\tablenotetext{b}{The masses are derived by assuming GR is valid.}
\tablecomments{
Figures in parentheses represent 1$\sigma$ uncertainties in the last quoted digit. Values from:    
$^1$\citealt{Kramer:2006aa}; 
$^2$\citealt{Fonseca:2014aa,Stairs:2002aa};
$^3$\citealt{Ferdman:2014aa}; 
$^4$\citealt{Weisberg:2016aa}; 
$^5$\citealt{Cameron:2018aa}; 
$^6$\citealt{Bhat:2008aa,Ord:2002aa}; 
$^7$\citealt{Freire:2012aa}; 
$^8$\citealt{Antoniadis:2013aa}; 
$^9$\citealt{Lazaridis:2009aa,Desvignes:2016aa};
$^{10}$\citealt{Desvignes:2016aa}.
}
\end{deluxetable*}
\end{longrotatetable}

\subsection{Method}\label{Method}
In SMG, the PK parameters \eqref{5_PK} contain the four unknown quantities ($m_p$, $m_c$, $\ep_p$, $\ep_c$) and the eight observables (${P}_b$, $e$, $x_p$, $\dot{\omega}$, $\gamma$, $r$, $s$, $\dot{{P}_b}$). In general, the latter can directly be obtained by analyzing the pulsar timing data. For the former, we can set up Monte Carlo simulations to constrain them. In the simulation, all observables are randomly sampled from a normal distribution with mean and standard deviation equal to their fitted values and uncertainties, respectively, and then these unknown quantities can be obtained by combining and solving Equations \eqref{5_PK}. This process is repeated $10^6$ times, and then we can construct the histograms for these unknown quantities to constrain them.

\section{Results and Discussion}\label{section5}
In this section, we derive the constraints on the scalar charges from the ten binary pulsar systems by performing the Monte Carlo simulations. The results are listed in Table \ref{tab2_eps}, and we discuss these in more detail below.

\begin{table}[!htbp]
\centering
\caption{Upper bound on scalar charge at 95.4\% CL for the ten binary pulsars\label{tab2_eps}}
\begin{tabular}{lr}
\hline\hline
NS$-$NS PSR    &  Upper bound on $\ep_{\N}$ \\
\hline
{J0737$-$3039} &  $6.6\times10^{-2}$ \\
{B1534$+$12}     &  $1.1\times10^{-1}$   \\
{J1756$-$2251}  &  $3.3\times10^{-1}$  \\
{B1913$+$16}     &   $3.3\times10^{-2}$ \\
{J1757$-$1854}  &  $1.7\times10^{-1}$   \\
\hline
NS$-$WD PSR  & Upper bound on $\ep_{\W}$ \\
{J1141$-$6545}   &  $4.3\times10^{-3}$ \\
{J1738$+$0333}  &  $3.4\times10^{-3}$ \\
{J0348$+$0432} &  $7.8\times10^{-3}$ \\
{J1012$+$5307}  &  $1.5\times10^{-2}$  \\
{J0751$+$1807}  &  $1.9\times10^{-2}$  \\
\hline
\end{tabular}
\end{table}

\subsection{NS-NS Binary Pulsars}
NS-NS binary pulsar is a symmetric system, we set $\ep_{p}=\ep_{c}=\ep_{\N}$, i.e., the scalar dipole radiation is neglected. In this system, the four unknown quantities reduce to three ($m_p$, $m_c$, $\ep_{\N}$), which are obtained by using the most accurate three observables in this subsection.

PSR J0737$-$3039 is the only known double pulsar, discovered in 2003 \citep{Burgay:2003aa}. It consists of a pulsar with a period of 22-ms, PSR J0737$-$3039A, in a 2.4-hour orbit with a younger pulsar with a period of 2.7-s, PSR J0737$-$3039B. Using the observables $\dot{\omega}$, $q$ and $s$ (see Table \ref{tab1_PSRs_param}), performing Monte Carlo simulations described in subsection \ref{Method}, we obtain the pulsar and companion masses of $m_{p}=(1.338\pm0.004)M_{\sun}$ and $m_{c}=(1.249\pm0.002)M_{\sun}$ in agreement with those of GR (see Table \ref{tab1_PSRs_param}), and a bound of $\ep_{\N}\leq0.066$ at 95.4\% CL. The masses imply $\dot{P}_b=(-1.249\pm0.006)\times10^{-12}$, $\gamma=(0.3842\pm0.0006)\rm ms$ and $r=(6.154\pm0.006)\rm{\mu}s$, which agree well with their observed values (see Table \ref{tab1_PSRs_param}), respectively.

PSR B1534$+$12 is a 37.9-ms radio pulsar in a 10.1-hour quasi-elliptical orbit with a NS companion. Using the PK parameters $\dot{\omega}$, $\gamma$ and $s$, we obtain the pulsar and companion masses of $m_{p}=(1.339\pm0.009)M_{\sun}$ and $m_{c}=(1.3458\pm0.0006)M_{\sun}$ in agreement with those of GR (see Table \ref{tab1_PSRs_param}), and a bound of $\ep_{\N}\leq0.11$ at 95.4\% CL. The masses imply $r=(6.629\pm0.003)\rm{\mu}s$ in good agreement with its observed value of $(6.6\pm0.2)\rm{\mu}s$.

PSR J1756$-$2251 is a 28.4-ms pulsar in a 7.67-hour quasi-elliptical orbit with a low-mass NS companion. Using the PK parameters $\dot{\omega}$, $\gamma$ and $\dot{P}_{b}$, we obtain the pulsar and companion masses of $m_{p}=(1.40\pm0.07)M_{\sun}$ and $m_{c}=(1.24\pm0.02)M_{\sun}$ in agreement with those of GR (see Table \ref{tab1_PSRs_param}), and a bound of $\ep_{\N}\leq0.33$ at 95.4\% CL. The masses imply $s=0.92\pm0.02$ and $r=(6.1\pm0.1)\rm{\mu}s$, which agree with their observed values (see Table \ref{tab1_PSRs_param}), respectively.

PSR B1913$+$16 was the first binary pulsar discovered \citep{Hulse:1975aa}. It consists of two NSs (one is an observed pulsar) orbiting in a very tight, highly elliptical orbit. This system only provides the three PK parameters $\dot\omega$, $\gamma$ and $\dot{P}_b$, which give the pulsar and companion masses of $m_{p}=(1.436\pm0.004)M_{\sun}$ and $m_{c}=(1.391\pm0.002)M_{\sun}$ and a bound of $\ep_{\N}\leq0.033$ at 95.4\% CL. The masses are good agreement with those of GR (see Table \ref{tab1_PSRs_param}).

PSR J1757$-$1854 is a 21.5-ms pulsar in a highly-elliptical, 4.4-hour orbit with a NS companion. Using the PK parameters $\dot{\omega}$, $\gamma$ and $\varsigma$, we obtain the pulsar and companion masses of $m_{p}=(1.32\pm0.06)M_{\sun}$ and $m_{c}=(1.392\pm0.007)M_{\sun}$ in agreement with those of GR (see Table \ref{tab1_PSRs_param}), and a bound of $\ep_{\N}\leq0.17$ at 95.4\% CL. The masses imply $\dot{P}_b=(-5.2\pm0.4)\times10^{-12}$ and $h_3=(5.0\pm0.9)\rm{\mu}s$, which agree with their observed values (see Table \ref{tab1_PSRs_param}), respectively.

\begin{figure*}[htbp]
\centering
\subfigure[PSR J0737$-$3039]{
\label{fig1} 
\includegraphics[width=5.8cm, height=5.8cm]{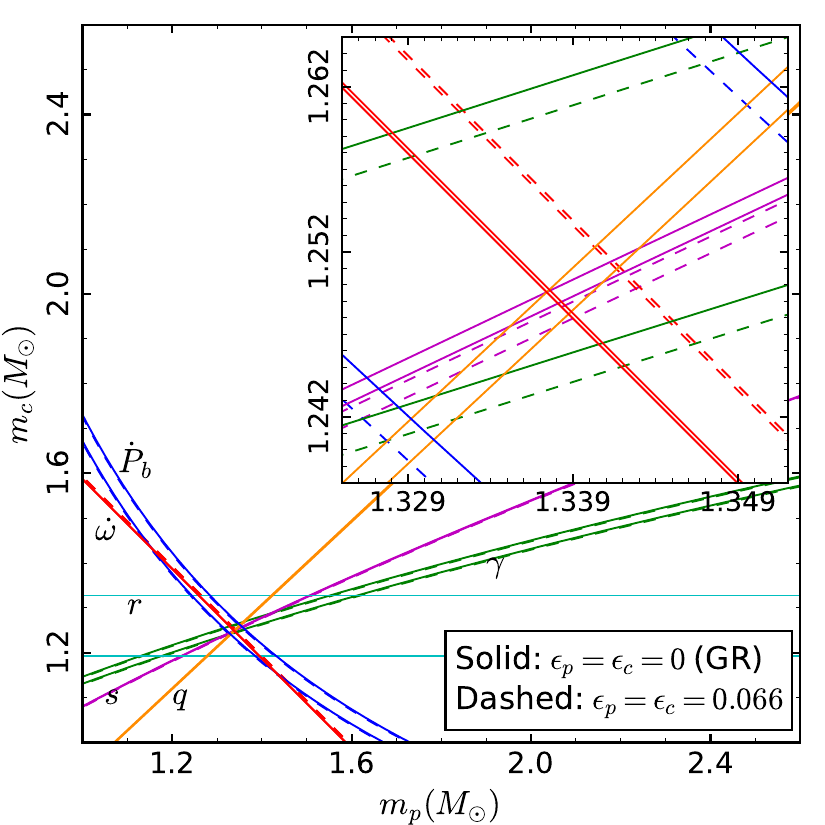}}
\subfigure[PSR B1534$+$12]{
\label{fig2} 
\includegraphics[width=5.8cm, height=5.8cm]{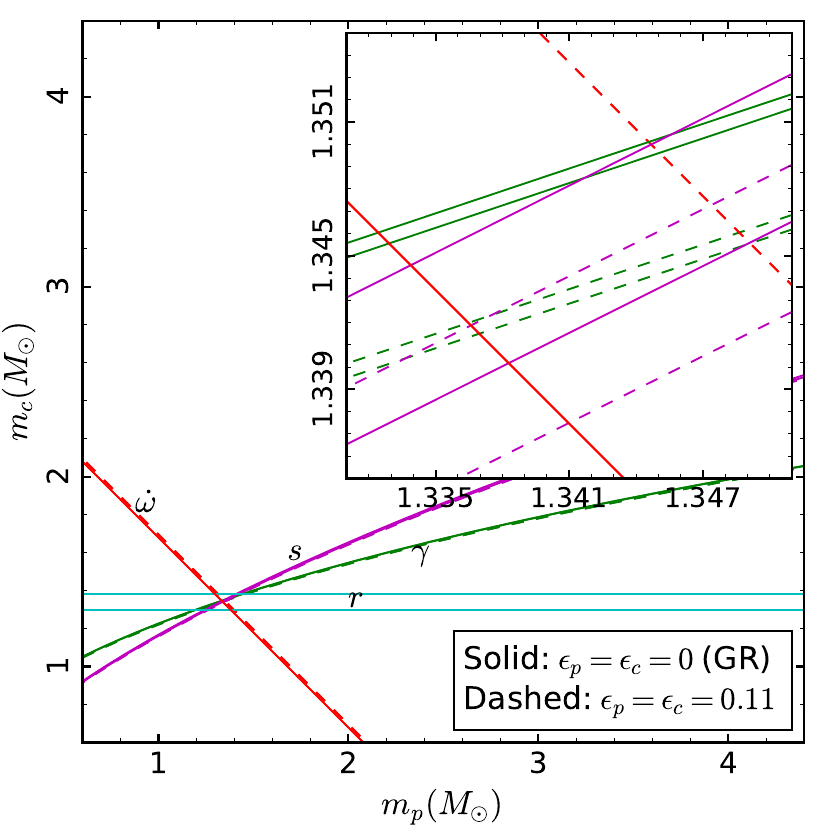}}
\subfigure[PSR J1756$-$2251]{
\label{fig3} 
\includegraphics[width=5.8cm, height=5.8cm]{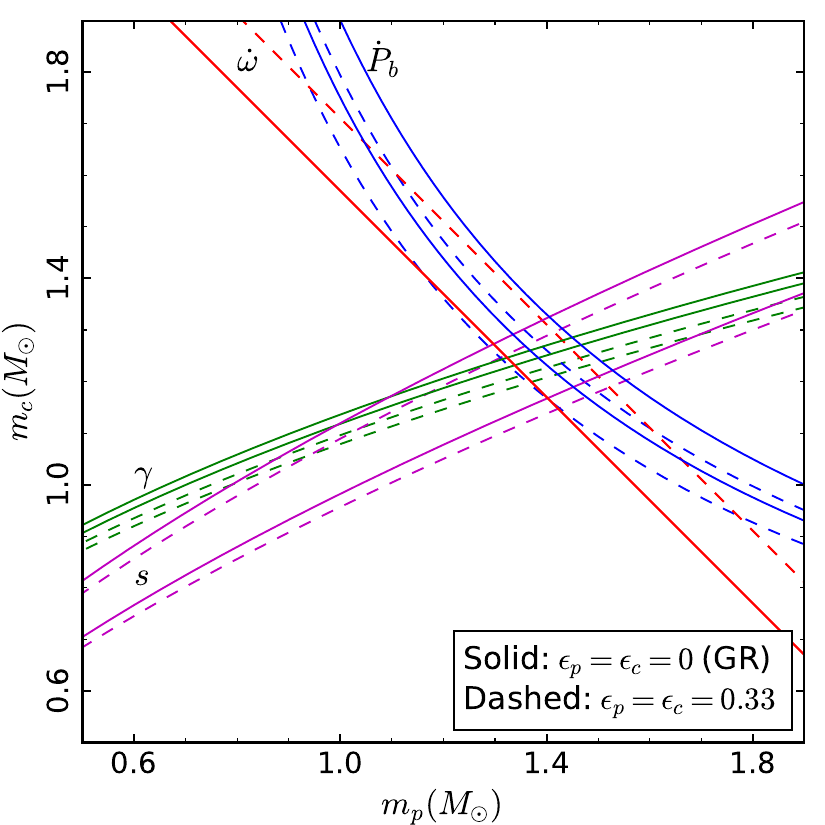}}
\subfigure[PSR B1913$+$16]{
\label{fig4} 
\includegraphics[width=5.8cm, height=5.8cm]{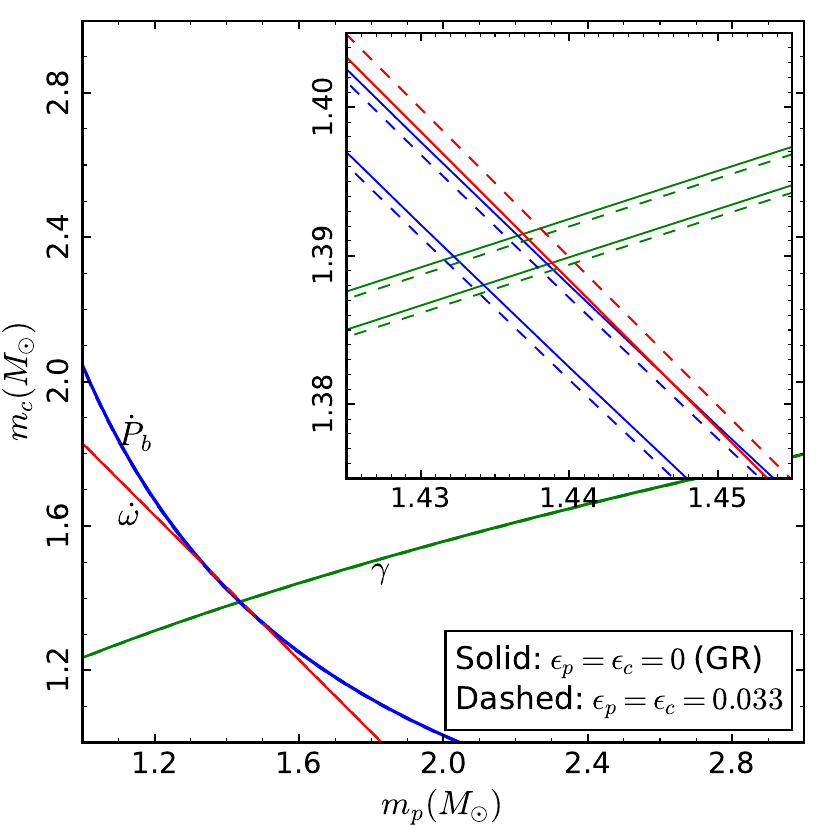}}
\hspace{30mm}
\subfigure[PSR J1757$-$1854]{
\label{fig5} 
\includegraphics[width=5.8cm, height=5.8cm]{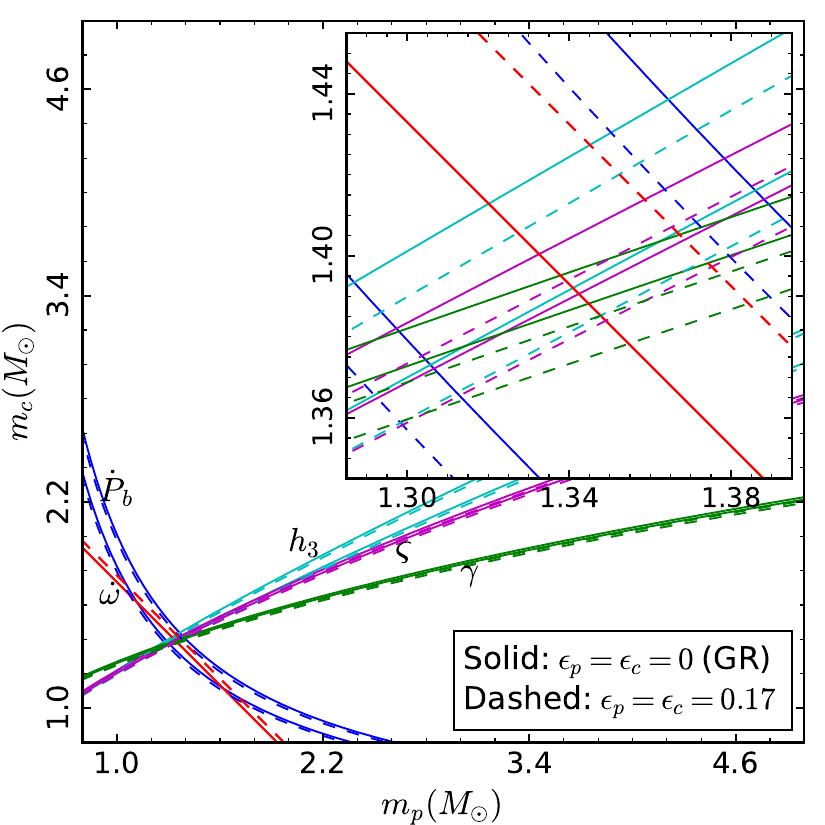}}
\caption{Mass-mass diagrams for the five NS-NS PSRs J0737$-$3039, B1534$+$12, J1756$-$2251, B1913$+$16 and J1757$-$1854. The constraints on $m_p$ and $m_c$ in GR (solid) and in SMG (dashed) are based on the PK parameters. The width of each curve represents $\pm1\sigma$ error bounds.}
\label{figNS_NS}
\end{figure*}

The best bound of $\ep_{\N}\leq0.033$ from PSR~B1913$+$16 excludes significant strong-field deviations from GR ($\ep_{\N}=0$). We construct the mass-mass diagrams and the constraints on $\ep_{\N}$ as shown in Figure \ref{figNS_NS}. The constraints on $m_p$ and $m_c$ in GR (solid lines) and in SMG (dashed lines) are based on the PK parameters. The results of all tests agree well with GR.

\subsection{NS-WD Binary Pulsars}
Now we consider NS-WD binary pulsar systems. The scalar charge is inversely proportional to the compactness (see Equation~\eqref{epsilon_a}), i.e., $\ep_{\W}/\ep_{\N}\simeq\Phi_{\N}/\Phi_{\W}\sim10^{4}$. That is, comparing to the WD scalar charge $\ep_{\W}$, the NS scalar charge $\ep_{\N}$ is approximately equal to zero. Therefore, the expressions \eqref{5_PK} of the PK parameters $\dot{\omega}$, $\gamma$, $r$ and $s$ reduce to those of GR, which leads that the masses $m_{\N}$ and $m_{\W}$ in SMG are the same as those in GR (see Table \ref{tab1_PSRs_param}). In the case of $\ep_{\N}\simeq0$, the orbital period decay is dominated by the dipole radiation (see Equation \eqref{P_decay2}),
\begin{align}
\dot{P}_b^{\rm dipole}=-\f{2\pi^2G}{P_b}\f{m_{\N}m_{\W}}{m_{\N}+m_{\W}}\f{(1+\f{e^2}{2})}{(1-e^2)^{5/2}}\ep_d^2,
\end{align}
where $\ep_d=\ep_{\W}-\ep_{\N}\simeq\ep_{\W}$. The constraints on $\ep_{\W}$ (or dipole radiation) only come from the orbital period decay rate $\dot{{P}_b}$.

\begin{figure*}[htbp]
\centering
\subfigure[PSR J1141$-$6545]{
\label{fig6} 
\includegraphics[width=5.8cm, height=5.8cm]{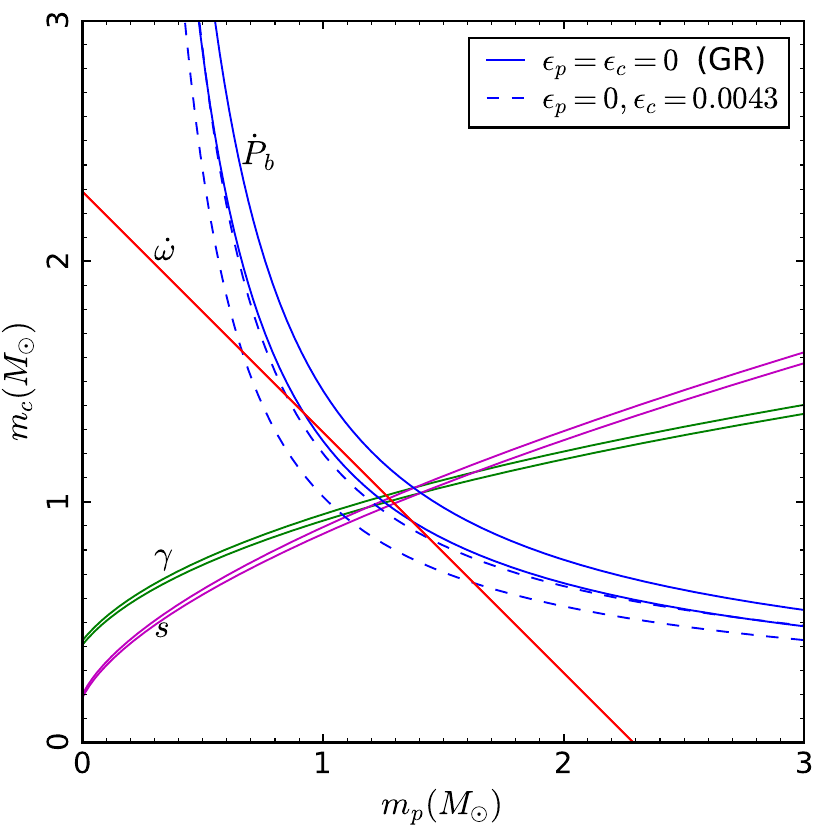}}
\subfigure[PSR J1738$+$0333]{
\label{fig7} 
\includegraphics[width=5.8cm, height=5.8cm]{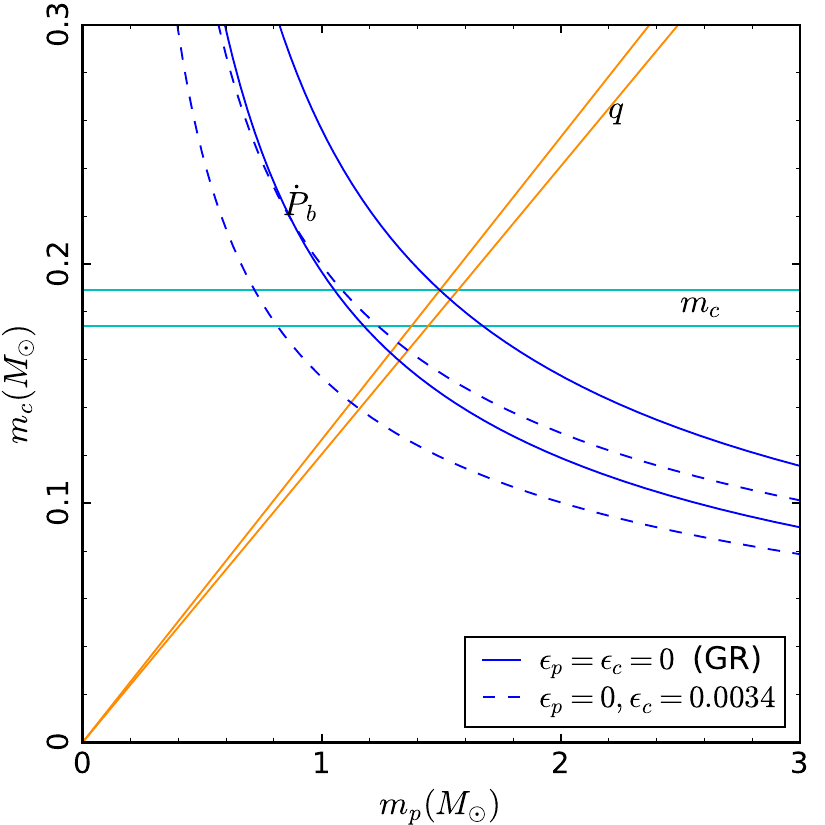}}
\subfigure[PSR J0348$+$0432]{
\label{fig8} 
\includegraphics[width=5.8cm, height=5.8cm]{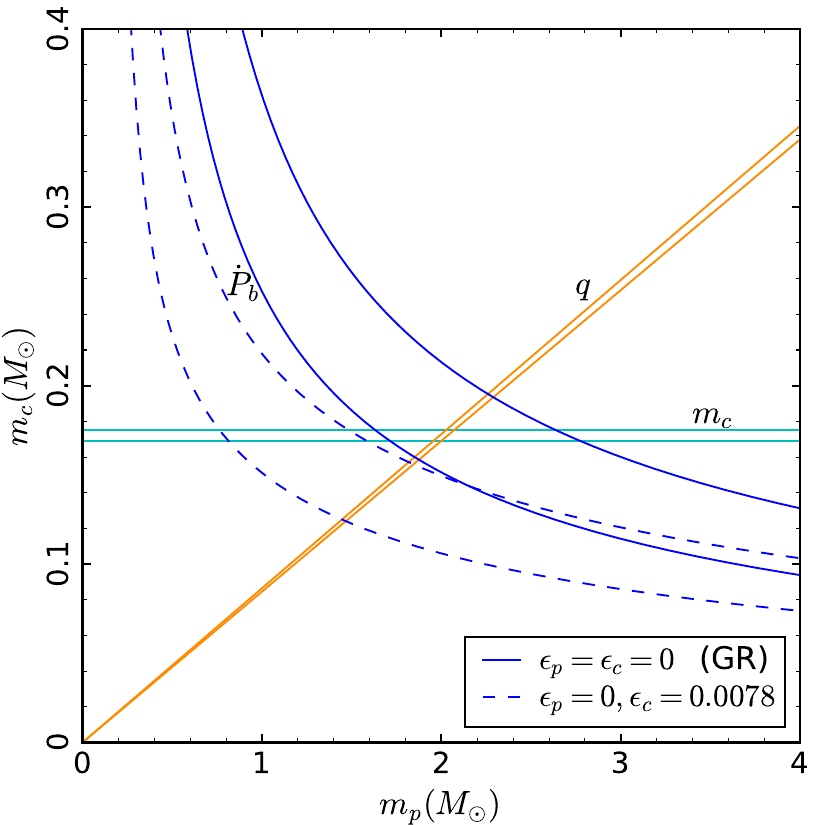}}
\subfigure[PSR J1012$+$5307]{
\label{fig9} 
\includegraphics[width=5.8cm, height=5.8cm]{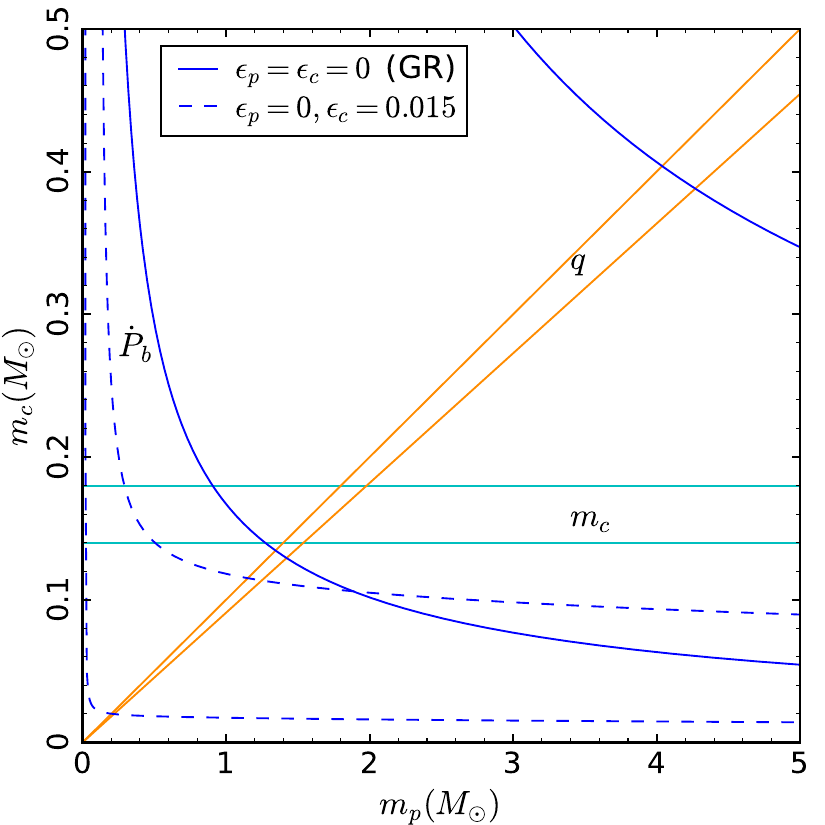}}
\hspace{30mm}
\subfigure[PSR J0751$+$1807]{
\label{fig10} 
\includegraphics[width=5.8cm, height=5.8cm]{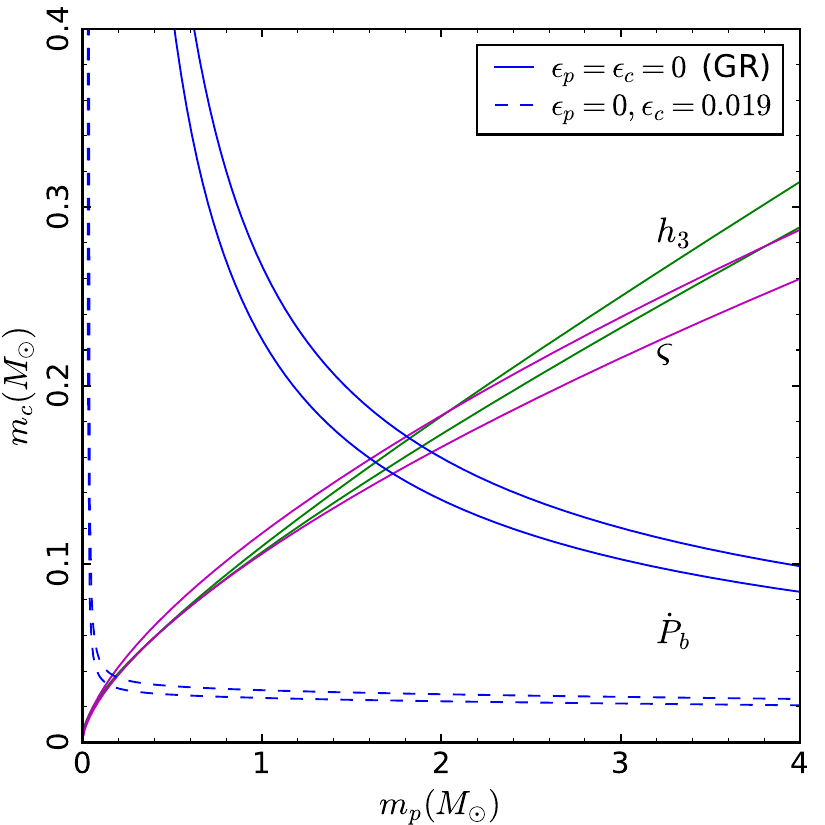}}
\caption{Mass-mass diagrams for the five NS-WD PSRs J1141$-$6545, J1738$+$0333, J0348$+$0432, J1012$+$5307 and J0751$+$1807. The constraints on $m_p$ and $m_c$ in GR (solid) and in SMG (dashed) are based on the PK parameters and the other observables. The width of each curve represents $\pm1\sigma$ error bounds.}
\label{figNS_WD}
\end{figure*}

PSR J1141$-$6545 is a pulsar in a quasi-elliptical orbit with a WD companion. PSRs J1738$+$0333, J0348$+$0432, J1012$+$5307 and J0751$+$1807 are MSPs in quasi-circular orbits with low-mass WD companions. Using the intrinsic $\dot{P}_b$ from the five systems, we obtain the upper bounds on $\ep_{\W}$ of $0.0043$, $0.0034$, $0.0078$, $0.015$ and $0.019$ (95.4\% CL), respectively. These exclude significant strong-field deviations and show good agreement with GR ($\ep_{\W}=0$). PSR~J1738$+$0333 provides the best bound of $\ep_{\W}\leq0.0034$. We find that the bounds on $\ep_{\W}$ are better than the bounds on $\ep_{\N}$ (see Table \ref{tab2_eps}), because of the dipole radiation in NS-WD binaries. The mass-mass diagrams and the constraints on $\ep_{\W}$ are shown in Figure \ref{figNS_WD}. The constraints on $m_{\N}$ and $m_{\W}$ in GR (solid lines) and in SMG (dashed lines) are based on the PK parameters and the other observables.

\subsection{Constraints on Scalar Background}
In this subsection, we derive the constraints on the scalar background $\ph$. For compact objects, the scalar charge can be approximated by $\ep=\ph/(\Mpl\Phi)$, where $\Phi=Gm/R$ is the compactness of the star and $R$ is its radius. In these systems under consideration, only two PSRs J0348$+$0432$^1$ and J1738$+$0333$^2$ (labeled by 1 and 2) provide the measured radii for WDs, i.e., $R_{\W1}=0.065^{+0.005}_{-0.005}\,R_{\sun}$ \citep{Antoniadis:2013aa} and $R_{\W2}=0.037_{-0.003}^{+0.004}\,R_{\sun}$ \citep{Freire:2012aa}, and we have $\Phi_{\W1}\!=\!5.61\!\times\!10^{-6}$ and $\Phi_{\W2}\!=\!1.04\!\times\!10^{-5}$. Using these and the corresponding bounds on $\ep_{\W}$, we obtain a bound on $\ph$ (95.4\% CL),
\begin{align}
\label{phi_m_1}
\f{\ph}{\Mpl}\leq4.4\times10^{-8}{\rm ~~~from~~J0348\!+\!0432},
\end{align}
and a better one 
\begin{align}
\label{phi_m_2}
\f{\ph}{\Mpl}\leq3.4\times10^{-8}{\rm ~~~from~~J1738\!+\!0333},
\end{align}
which agrees well with the upper bound of $3.3\times10^{-8}$ in the previous work \citep{Zhang:2017aa}. This result is applicable to the general SMG and has been used to place the constraints on chameleon, symmetron, dilaton \citep{Zhang:2017aa} and $f(R)$ \citep{Liu:2018aa} theories.

\section{conclusions}\label{section6}
As a simple generalization of GR, SMG is a class of scalar-tensor gravity with screening mechanism in order to satisfy the tight tests from the Solar System. Binary pulsars are the ideal laboratories for testing gravity. In this paper, we investigated how the screening effect affects the PK parameters and placed the tight constraints on the theory by the pulsar observations.

We first studied the orbital dynamics of the binary pulsars and then calculated in detail the five PK parameters in this theory. The PK parameters depend on the masses and scalar charges of the system. The scalar charge characters the difference from GR and is inversely correlated to the compactness in this theory. Thus, the deviations from GR become small for compact objects, which is very different from most alternative theories of gravity without screening mechanisms.
We placed the constraints on the scalar charges by using the PK observables from the ten binary pulsars. The dipole radiation in NS-WD binaries leads to more stringent constraints on deviations from GR. PSR~B1913$+$16 is the most constraining system for NS scalar charge. PSR~J1738$+$0333 is the most constraining system for WD scalar charge and gives the best constraint on the scalar background. The results of all tests rule out significant strong-field deviations and agree well with GR.

\acknowledgments{\it Acknowledgements} We appreciate the helpful discussion with Shenghua Yu, Zesen Lin, and Yulong Gao. This work is supported by NSFC No. 11773028, 11603020, 11633001, 11173021, 11322324, 11653002, 11421303, project of Knowledge Innovation Program of Chinese Academy of Science, the Fundamental Research Funds for the Central Universities and the Strategic Priority Research Program of the Chinese Academy of Sciences Grant No. XDB23010200.

\newpage

\begin{appendix}
\section{PN expansion of the scalar field}\label{appendix_a}
We solve the scalar field equation by the method of matching the internal and external solutions. Now, considering a static spherically symmetric source object with constant density $\rho_o$ and radius $R$, it is embedded in a homogeneous background of matter density $\rho_b$. Then, the scalar field equation \eqref{scalar_eom} can reduce to
\be
\label{scalar field eq.3}
\frac{{\rm d}^2\phi}{{\rm d}r^2}+\frac 2r\frac{{\rm d}\phi}{{\rm d}r}=m^2_{\rm m}(\rho)\big[\phi-\phi_{\rm m}(\rho)\big],
\ee
with
\begin{equation}
\rho(r) = \left\{
\begin{matrix}
\rho_{o}\qquad~{\rm for}~~~ r<R \cr
\rho_{b}\qquad~ {\rm for}~~~ r>R
\end{matrix}
\right..
\end{equation}
This is a second order differential equation, so we must impose two boundary conditions: ${\rm d\phi}/{\rm d}r\big|_{r=0} = 0$ and $\phi\big|_{r\rightarrow\infty}\rightarrow\ph$. Moreover, the scalar field $\phi$ and ${\rm d}\phi/{\rm d}r$ are of course continuous at the boundary. By solving Equation \eqref{scalar field eq.3} directly, yields the exact solution
\begin{subequations}
\begin{align}
\phi(r<R)&=\phi_o+\frac Ar\sinh(m_or),
\\
\label{exterior scalar field}
\phi(r>R)&=\ph+\frac Bre^{-m_s r},
\end{align}
\end{subequations}
with
\begin{subequations}
\begin{align}
A&=\frac{(\ph-\phi_o)(1+m_s R)}{m_o\cosh(m_oR)+m_s\sinh(m_oR)},
\\
B&=-e^{m_s R}(\ph-\phi_o)\frac{m_oR-\tanh(m_oR)}{m_o+m_s\tanh(m_oR)},
\end{align}
\end{subequations}
where $\phi_o$ and $\ph$ are the positions of the effective potential minimum inside and far outside the source object, respectively, and $m_o$ and $m_s$ are the scalar field masses at $\phi_o$ and $\ph$, respectively.
In the case of $m_o^{-1}\ll R\ll m_s^{-1}$, the exterior scalar field \eqref{exterior scalar field} reduces to
\begin{align}\label{exterior scalar solution}
\begin{split}
\varphi(r)&=\phi(r)-\ph=- \Mpl\frac{Gm\epsilon}re^{-m_s r},
\end{split}
\end{align}
with
\be\label{epsilon}
\epsilon\equiv\frac{\ph-\phi_o}{\Mpl\Phi},
\ee
where $m$ is the mass of the object, $\Phi=Gm/R$ is its compactness, and $\epsilon$ is its scalar charge (or screened parameter).
For a $N$-body system, the scalar field is given by
\begin{align}
\varphi=\sum_a\varphi_a=-\Mpl\sum_a\frac{Gm_a\ep_a}{r_a}e^{-m_sr_a}\label{multibody_varphi},
\end{align}
where $r_a=\left|\mathbf{r}-\mathbf{r}_a(t)\right|$.

Now let us solve the scalar field equation by using the PN approximation \citep{Will:1993aa,Will:2014aa}. Up to Newtonian order, the field equation \eqref{linear_scalar_eq} reduces to
\begin{align}\label{t_ind_sca_equ}
\left(\nabla^2-m^2_s\right)\ac{(2)}{\varphi}=\ph^{-1}\sum_as_am_a\delta^3\big(\mathbf{r}-\mathbf{r}_a(t)\big),
\end{align}
and using Green's function method the solution is
\begin{align}\label{sen_sca_eq_2}
\ac{(2)}{\varphi}=-2\f{\Mpl^2}{\ph}\sum_a\frac{Gm_as_a}{r_a}e^{-m_sr_a}.
\end{align}
By comparing the above two solutions \eqref{multibody_varphi} with \eqref{sen_sca_eq_2}, yields the useful relation between sensitivity and scalar charge,
\begin{align}\label{s_a}
s_a=\frac{\ph}{2\Mpl}\ep_a.
\end{align}

In the near zone, neglecting the scalar field mass $m_s$ of cosmological scales, using Equations \eqref{sen_sca_eq_2} and \eqref{PN_tensor_ij}, the scalar field equation \eqref{linear_scalar_eq} reduces to
\begin{align}\label{phi24}
\begin{split}
\square{\varphi}=&4\pi\Mpl\sum_aGm_a\ep_a\delta^3\big(\mathbf{r}-\mathbf{r}_a(t)\big)\!\times\!\bigg[1-\f12v_a^2-\sum_{b\ne a}\f{Gm_b}{r_{b}}-\f{s'_a}{s_a}\f{\Mpl}{\ph}\sum_{b\ne a}\f{Gm_b\ep_b}{r_{b}}\bigg]+\mathcal{O}(v^6),
\end{split}
\end{align}
where we have neglected the terms $V_n$, since the effects of dark energy are very weak in the near zone. The equation can be solved by using Green's function method,
\begin{align}
\square G_a(t,\mathbf{r})=-4\pi \delta^3(\mathbf{r}-\mathbf{r}_a(t))
\end{align}
with 
\begin{align}
 G_a(t,\mathbf{r})=\f{1}{|\mathbf{r}-\mathbf{r}_a(t)|}+\f12\f{\partial^2}{\partial t^2}|\mathbf{r}-\mathbf{r}_a(t)|+\f{1}{r_a}\mathcal{O}(v^4).
\end{align}
Using this, the solution of Equation \eqref{phi24} is presented in Equation \eqref{PN_scalar}.

\end{appendix}


\end{document}